\documentclass{llncs}
\usepackage[utf8]{inputenc}
\usepackage{amssymb,amsmath}
\usepackage{bm}
\usepackage{sistyle}
\usepackage{caption}
\usepackage{acronym}
\usepackage{graphicx}
\usepackage{afterpage}
\usepackage[colorlinks=true, allcolors=blue]{hyperref}
\usepackage{url}
\usepackage[capitalise]{cleveref}
\usepackage{multibib}
\usepackage[retainorgcmds]{IEEEtrantools}

\newacro{pdf}{probability density function}
\newacro{TV}{Total Variation}
\newacro{KDE}{Kernel Density Estimation}
\newacro{FWHM}{Full Width at Half Maximum}
\newacro{fMRI}{Functional Magnetic Resonance Imaging}
\newacro{FFT}{Fast Fourier Transform}
\newacro{ALE}{Activation Likelihood Estimation}
\newacro{VBM}{Voxel Based Morphometry}
\newacro{TFIDF}{Term Frequency - Inverse Document Frequency}

\DeclareMathOperator*{\tv}{TV}
\DeclareMathOperator*{\argmin}{argmin}
\DeclareMathOperator*{\argmax}{argmax}
\DeclareMathOperator*{\Tr}{Tr}

\DeclareMathOperator*{\diag}{diag}
\DeclareMathOperator*{\supp}{supp}

\renewcommand{\paragraph}[1]{\vspace{5pt}\noindent\textbf{#1}}

\newcites{annex}{References}
\begin{document}
\frontmatter
\mainmatter
\title{Text to brain:  predicting the spatial distribution of neuroimaging observations from text reports}








\author{Jérôme Dockès$^1$\and Demian Wassermann$^1$ \and Russell Poldrack$^2$
  \and Fabian Suchanek$^3$ \and Bertrand Thirion$^1$\and Gaël Varoquaux$^1$}

\institute{$^1$INRIA, CEA, Université Paris-Saclay, $^2$Stanford University,
$^3$Télécom ParisTech}

\maketitle              
\begin{abstract}
Despite the digital nature of magnetic resonance imaging, 
the resulting observations are most frequently
reported and stored in text documents. There is a trove of information untapped in medical health records, case reports, and medical publications. In this paper, we propose to mine brain medical publications to learn the
spatial distribution associated with anatomical terms. The problem is formulated
in terms of minimization of a risk on distributions which leads to a
least-deviation cost function. An efficient algorithm in the dual then learns
the mapping from documents to brain structures. 
Empirical results using coordinates extracted from the
brain-imaging literature show that i) models must adapt to semantic variation in
the terms used to describe a given anatomical structure, ii) voxel-wise
parameterization leads to higher likelihood of locations reported in unseen
documents, iii) least-deviation cost outperforms least-square. As a proof of concept for our method, we use our model of spatial distributions to predict the distribution of specific neurological conditions from text-only reports.

\end{abstract}
\section{Introduction}
Hundreds of thousands of studies, case reports, or patient records, capture 
observations in human neuroscience, basic or clinical.
Statistical analysis of this large amount of data could provide new insights.
Unfortunately, most of the spatial information that these data contain is difficult to
extract \emph{automatically}, because it is hidden in unstructured text, in
sentences such as:
``[...] in the \underline{anterolateral temporal
cortex}, especially the \underline{temporal pole} and \underline{inferior} and \underline{middle temporal gyri}''
\cite{mummery2000voxel}.
%
%
%
%

This data cannot be processed easily by a machine, as a machine does not know
where the temporal cortex is. 
As we will show, simply looking up such terms in atlases does not
suffice.
Indeed, even atlases
disagree \cite{bohland2009brain}. Furthermore, joint processing of many 
reports faces varying terminologies, with regions represented in different atlases that differ and overlap. Finally, not all terms in a report carry the same importance, and
practitioners use terms that are not the exact labels of any
atlas. Coordinate-based meta-analyses capture the spatial distribution of
a term from the literature
\cite{laird2005ale,yarkoni2011large}, but they also lack a model to combine terms.

Here, we propose to map case reports
automatically to the brain locations that they discuss: we learn mappings of anatomical terms to brain regions from medical publications.
%
%
We propose a new learning framework  for translating anatomical terms to brain images -- a process that we call ``encoding''.
We learn such a mapping, quantify its performance, and compare
possible choices of representation of spatial data. We then show in a
proof of concept that our model can predict the brain area for textual case reports.

\newcommand{\ignore}[1]{}
\ignore{
While such reports discuss brain structures and loci, systematic 
statistical analysis of the relevant positions and their spatial
variability is challenging.
On the opposite, medical images such as segmentations of the observed
structures lend themselves well to analysis. To exploit the information
in reports, we propose to translate the text into
brain maps or distributions over the brain.

However, relating text to brain areas calls for more than simple
heuristics.
Given the name of a brain region, an
atlas may provide its precise location, but even atlases
disagree \cite{bohland2009brain}. In addition, we wish to jointly process many reports, that mention regions from different terminologies, represented in different
atlases that differ and overlap. Further, not all terms inside a report carry the same importance.
Practitioners use terms that are not the exact labels of any
given atlas. How do we associate a brain image with each of these reports? As we
will see, when we make the jump from the controlled vocabulary of atlas labels
to free text, rigid rules fail and statistical models of the link
between terms and positions in the brain become necessary.
Here, we propose a formal framework to learn to translate
neuroscience text into brain images - a process that we call "encoding". For the
first time, we learn such a mapping, quantify its performance, and compare
possible choices of representations of spatial data.
}

\section{Methods: formalizing text-to-brain-map translation}\label{sec:methods}

\subsection{Problem setting: from text to spatial distributions}

We want to predict the likelihood of the location of relevant brain
structures described in a document. For this purpose, we perform supervised
learning on a corpus of brain-imaging studies, each containing: (i) a
text, and (ii) the locations -- \textit{i.e.} the stereotactic coordinates -- of its
observations. Indeed, \ac{fMRI} studies report the
coordinates of  activation peaks (\textit{e.g.,} \cite[Table 1]{van2011sound}), and
\ac{VBM} analyses report the location of differences
in gray matter density (\textit{e.g.,} \cite[Table 2]{mummery2000voxel}).
Following neuroimaging meta-analyses \cite{laird2005ale}, we frame the
problem in terms of spatial distributions of observations in the brain.
In a document, observed locations 
$\mathcal{L} = \{l_a \in \mathbb{R}^3, a = 1 \dots c\}$
are sampled from a \ac{pdf}~$p$ over the brain. \textbf{Our goal is to 
predict this \ac{pdf} $p$ from
the text $\mathcal{T}$}. We denote $q$ our predicted \ac{pdf}. A
predicted \ac{pdf} $q$ should be close to $p$, or take high values at the
coordinates actually reported in the study:
$\prod_{l \in \mathcal{L}} q(l)$ must be large.
In a supervised learning setting, we start from a collection of studies
$\mathcal{S} = (\mathcal{T}, \mathcal{L})$, with $\mathcal{T}$ the text
and $\mathcal{L}$ the locations. Building the prediction engine then
entails the choice of a model relating the predicted \ac{pdf} $p$ to the
text $\mathcal{T}$, the choice of a loss, or data-fit term, and
some regularization on the model parameters.
We now detail how we make each of these choices to construct a prediction.

\paragraph{Model.}
We start by modelling the dependency of our spatial \ac{pdf} $q$ on the
study text $\mathcal{T}$. This entails both choosing a representation for
$q$ and writing it as a function of the text. While $q$ is defined on a
subvolume of $\mathbb{R}^3$, the brain volume, we build it using a
partition to work on a finite probably space: this can be either a
regular grid of voxels or a set of anatomical regions (i.e. an atlas)
$\mathcal{R} = \{\mathcal{R}_k, k=0\dots m\}$.
As such a partitioning imposes on each region to be homogeneous, $q$ is then
formally written on $\mathbb{R}^3$ in terms of the indicator functions of
the parts\footnote{$\mathcal{R}_0$  denotes the volume
outside of the brain, or background, on which $q$ is 0.}: $\{r_k = \frac{\mathbb{I}_k}{\|\mathbb{I}_k\|_1}, ~ k = 1 \dots m\}$.
Importantly, the volume of each part $\|\mathbb{I}_k\|_1$ appears as
a normalization constant.

To link $q$ to the text $\mathcal{T}$ of the study, we start
by building a term-frequency vector representation of $\mathcal{T}$, which we denote
$\bm{x} \in \mathbb{R}^d$. $d$ is the size of our vocabulary of English words
$\mathcal{W} = \{w_t\}$, and $\bm{x}_t$ is the frequency of word $w_t$ in the text.
We assign to each atlas region a weight that depends
linearly on $\bm{x}$:
\begin{equation}
  q(z) = \sum_{t=1}^d \sum_{k=1}^m \bm{x}_t \bm{\beta}_{t,k}r_k(z) \quad \forall z \in \mathbb{R}^3
    \label{eq:search-space}
\end{equation}
where $\bm{\beta} \in \mathbb{R}^{d \times m}$ are model parameters, which we
will learn.

Using an atlas is a form of regularization: constraining the
prediction to be in the span of $\{r_k\}$ reduces the size of the
search space. Fine partitions, \emph{e.g.} atlases with many regions or
voxel grids, yield models with more expressive power, but more likely
to overfit. Choosing an atlas thus amounts to a bias-variance
tradeoff.

\paragraph{Label-constrained encoder.}
A simple heuristic to turn a text into a brain map is to
use atlas labels and ignore interactions between terms.
The probability of a region is taken to be proportional to the frequency
of its label in the text.
The vocabulary is then the set of labels: $d = m$.
As the word $w_k$ is the label of $\mathcal{R}_k$, $\bm{\beta}$ is diagonal.
For example, for a region $\mathcal{R}_k$ in the atlas labelled ``parietal lobe'',
the probability on $\mathcal{R}_k$
depends only on the frequency of the phrase ``parietal lobe'' in the text.
We call this model \emph{label-constrained encoder}.

\subsection{Loss function: measuring errors on spatial distributions}

\paragraph{Strategy.}
We will fit the coefficients $\bm{\beta}$ of our model, see
\cref{eq:search-space}, by minimizing a risk $\mathcal{E}(p, q)$:
the expectation of a distance between $p$ and
%
%
$q$.

\paragraph{A plugin estimator of $p$.}
We do not have access to the true \ac{pdf}, $p$; we need a plugin estimator,
which we denote $\hat{p}$.
%
%
%
By construction of our prediction $q$, the best approximation of $p$ we can hope for belongs to
the span of our regions $\{r_k\}$. Hence, we build our estimator $\hat{p}$
in this space, setting the probability of a region to be proportional to the number of coordinates that fell inside it:
\begin{equation}
  \hat{p} = \sum_{k=1}^m \frac{|\{a, \mathbb{I}_k(l_a) = 1\}|}{c} r_k  ~ = ~
  \sum_{k=1}^m \frac{1}{c} \sum_{a=1}^c\mathbb{I}_k(l_a) r_k ~ \triangleq ~
   \sum_{k=1}^m \hat{\bm{y}}_k r_k
   \quad
   .
\end{equation}
When regions are voxels, there are too many regions and
too few coordinates. Hence we use Gaussian \ac{KDE} to 
smooth the estimated \ac{pdf}\footnote{Using an atlas is also a form of
  KDE, with kernel $(z, z') \mapsto 1 / \|\mathbb{I}_k\|_1$ if $z$ and $z'$ belong to
the same region $\mathcal{R}_k, k \in \{1,\dots m\}$, 0 otherwise.}.
Our supplementary material details a
fast \ac{KDE} implementation.
%

\paragraph{Choice of $\mathcal{E}$.}
We use two common distance functions for our loss.  The first is \ac{TV}, a common
distance for distributions. Note that $p$ defines a probability measure on
the finite sample space
$\mathcal{R}$, $\mathcal{P}(\mathcal{R}_k)= \int_{\mathcal{R}_k} p(z) dz$, where $\mathcal{R} =
\{\mathcal{R}_k, k=1 \ldots m\}$ and $\mathcal{R}_k = \supp(r_k)$. $q$ defines $Q$ in the same way. Then,
\begin{equation}
\tv(\mathcal{P}, \mathcal{Q}) = \sup_{\mathcal{A} \subset \mathcal{R}}|\mathcal{P}(\mathcal{A}) - \mathcal{Q}(\mathcal{A})| \quad .
\end{equation}
Since $\mathcal{R}$ is finite, a classical result (see
\cite{gibbs2002choosing}) shows that this supremum is attained by
taking $\mathcal{A} = \{\mathcal{R}_k | \mathcal{P}(\mathcal{R}_k) >
\mathcal{Q}(\mathcal{R}_k)\} $ (or its complementary) and:
%
%
\begin{equation}
  \tv(\mathcal{P}, \mathcal{Q}) = \frac{1}{2}\sum_{k=1}^m|\mathcal{P}(\mathcal{R}_k) - \mathcal{Q}(\mathcal{R}_k)| = \frac{1}{2}\int_{\mathbb{R}^3}|p(z) - q(z)| dz \quad .
\end{equation}
The \ac{TV} is half of the $\ell_1$ distance between the \ac{pdf}s.
$\|\hat{p} - q\|_1$ is therefore a natural choice for our loss. The second choice is
$\|\hat{p} - q\|_2^2$, which is a popular distance and has
the appeal of being differentiable everywhere.
%
%

\paragraph{Factorizing the loss.} Let us call $v_k$ the volume of $r_k$, i.e. the
size of its support:
$v_k \triangleq \|\mathbb{I}_k\|_1, ~ k = 1 \dots m$.
Remember that $r_k = \frac{1}{v_k}\mathbb{I}_k$.
Our loss can now  be factorized (see supplementary material for details):
\begin{IEEEeqnarray}{rCl}
   \int_{\mathbb{R}^3} \delta(\hat{p}(z) - q(z))dz & = &%
    \sum_{k=1}^m v_k \delta \left(\frac{\hat{\bm{y}}_k}{v_k} - \frac{\sum_{t=1}^d\bm{x}_{t}\bm{\beta}_{t,k}}{v_k} \right)
\end{IEEEeqnarray}
Here, $\delta$ is either the absolute value of the difference or the squared difference.
%
%

\subsection{Training the model: efficient minimization approaches}

To set the model parameters $\bm{\beta}$, we used $n$ example
studies $\{\mathcal{S}_i = (\mathcal{T}_i, \mathcal{L}_i), ~ i = 1 \dots n \}$.
We learn $\bm{\beta}$ by minimizing the empirical risk on $\{S_i\}$ and an $\ell_2$
penalty on $\bm{\beta}$.
%
%
%
We add to the previous notations the index~$i$ of each example:
$p_i$, $q_i$, $\hat{\bm{y}}_i$, $\bm{x}_i$. $\hat{\bm{Y}} \in
\mathbb{R}^{n \times m}$ is the
matrix such that $\hat{\bm{Y}}_{i} = \hat{\bm{y}}_i$, and $\bm{X} \in \mathbb{R}^{n \times d}$
such that $\bm{X}_i = \bm{x}_i$.
%

\paragraph{Case $\delta = \ell_2^2$.}
The empirical risk is
%
%
\begin{equation}
\sum_{i=1}^n \sum_{k=1}^m \left(\frac{\hat{\bm{Y}}_{i,k}}{\sqrt{v_k}} - \sum_{t=1}^d\frac{1}{\sqrt{v_k}}\bm{X}_{i,t}\bm{\beta}_{t,k} \right)^2.
\end{equation}
Defining $\bm{Y}'_{:,k} = \frac{\hat{\bm{Y}}_{:,k}}{\sqrt(v_k)}$ and
$\bm{\beta}'_{:,k} = \frac{\bm{\beta}_{:,k}}{\sqrt(v_k)}$, with an $\ell_2$
penalty, the problem is:
\begin{equation}
  \argmin_{\bm{\beta}'}\left(\|\bm{Y}' - \bm{\beta}'\bm{X}\|_2^2 + \lambda \|\bm{\beta}'\|_2^2 \right)
\end{equation}
where $\lambda \in \mathbb{R}_+$.
This is the least-squares ridge regression predicting $\hat{p}$ expressed in the
orthonormal basis of our search space $\{\frac{r_k}{\|r_k\|_2}\}$.

\paragraph{Case $\delta = \ell_1$.}
The empirical risk becomes
\begin{equation}
  \sum_{i=1}^n \sum_{k=1}^m|\hat{\bm{Y}}_{i,k} - \sum_{t=1}^d\bm{X}_{i,t}\bm{\beta}_{t,k}|
\end{equation}
This problem
is also known as a least-deviations regression, a particular case of quantile
regression \cite{koenker1978regression}, \cite{chen2005computational}.
Unlike $\ell_2$ regression,
which provides an estimate of the conditional mean of the target variable,
$\ell_1$ provides an estimate of the median.
 Quantile regression has been studied
(e.g. by economists), as it is more robust to outliers and better-suited than
least-squares when the noise is not normally distributed \cite{koenker1978regression}.
Adding an $\ell_2$ penalty, we have the minimization problem:
\begin{equation}
  \hat{\bm{\beta}} = \argmin_{\bm{\beta}}\left( \|\hat{\bm{Y}} - \bm{X}\bm{\beta}\|_1 + \lambda \|\bm{\beta}\|_2^2  \right)
    \label{eq:quantile_regression}
\end{equation}
%
%
Unpenalized quantile regression is often written as a linear
program and solved with the simplex algorithm
\cite{koenker1994remark}, iteratively reweighted least
squares, or interior point methods \cite{portnoy1997gaussian}. 
\cite{yi2017semismooth} uses a coordinate-descent to solve
a differentiable approximation of the quantile loss
(the Huber loss) with elastic-net penalty. Here, we minimize 
\cref{eq:quantile_regression} via its dual formulation (c.f. supplementary material):
%
%
%
\begin{align}
  \hat{\bm{\nu}} = \argmax_{\bm{\nu}}\left( Tr(\bm{\nu}^T\hat{\bm{Y}} -
\frac{1}{4\lambda}\bm{\nu}^T\bm{X}\bm{X}^T\bm{\nu}) \right)   \quad
   \text{ s.t. } \|\bm{\nu}\|_\infty \leq 1,
\end{align}
%
%
where $\bm{\nu} \in \mathbb{R}^{n \times m}$.
The primal solution is given by $\hat{\bm{\beta}} = \frac{\bm{X}^T\hat{\bm{\nu}}}{2\lambda}$.
As the dual loss $g$ is differentiable and the constraints are
\emph{bound} constraints, we can use an efficient quasi-Newton method
(L-BFGS,
\cite{byrd1995limited}).
%
%
%
$g$ and its gradient are fast to compute as $\bm{X}$ is sparse.
%
%
%
%
%
$\lambda$ is set by cross-validation on the training set.
We use warm-start on the regularization path (decreasing
values for $\lambda$) to initialize each problem.

\paragraph{Training the label-constrained encoder.} The columns of $\bm{\beta}$ can be fitted independently from
each other. If we want $\bm{\beta}$ to be diagonal, we only include one feature
in each regression: we fit $m$ univariate regressions $\bm{\hat{y}}_{:,k} \simeq
\bm{X}_{:,k}\bm{\beta}_{k,k}$.
%
%

\subsection{Evaluation: a natural model-comparison metric}

Our metric is the mean log-likelihood of an article's coordinates in the predicted
distribution, which diverges wherever $q=0$.
 we add a uniform background to the prediction, to ensure that it is non-zero everywhere:
%
%
\begin{flalign}
    \text{the predicted \ac{pdf} is written}
    &&
    q' = \frac{1}{2} (\sum_{k=1}^m\frac{\mathbb{I}_k}{v_k} + q)
    &&
\\
    \text{the score for a study $\mathcal{S}_i=(\mathcal{T}_i,
      \mathcal{L}_i), \mathcal{L}_i = \{l_{i,a}\}$ is}
    &&
  \frac{1}{c_i}\sum_{a=1}^{c_i}\log(q'_i(l_{i, a}))
    &&\label{eq:pseudo-ll}
\end{flalign}
%
%


\section{Empirical study}

\subsection{Data: mining neuroimaging publications}
We downloaded roughly 140K neuroimaging articles from online sources including
Pubmed Central and commercial publishers.
About 14K of these contain coordinates, which we extracted, as in
\cite{yarkoni2011large}.
We built a vocabulary of around 1000 anatomical region names by grouping
the labels of several atlases and the Wikipedia page ``List of regions in the human
brain''\footnote{\url{https://en.wikipedia.org/wiki/List_of_regions_in_the_human_brain}}. So in practice, $n \approx 14\cdot 10^3$ and $d
\approx 1000$. $m$ depends on the atlas (or voxel grid) and ranges from 20 to 30K.
%

\subsection{Text-to-brain encoding performance}\label{sec:encoding-performance}

\begin{figure}[t!]
\begin{minipage}[c]{0.67\textwidth}
  \includegraphics[width=1.\textwidth]{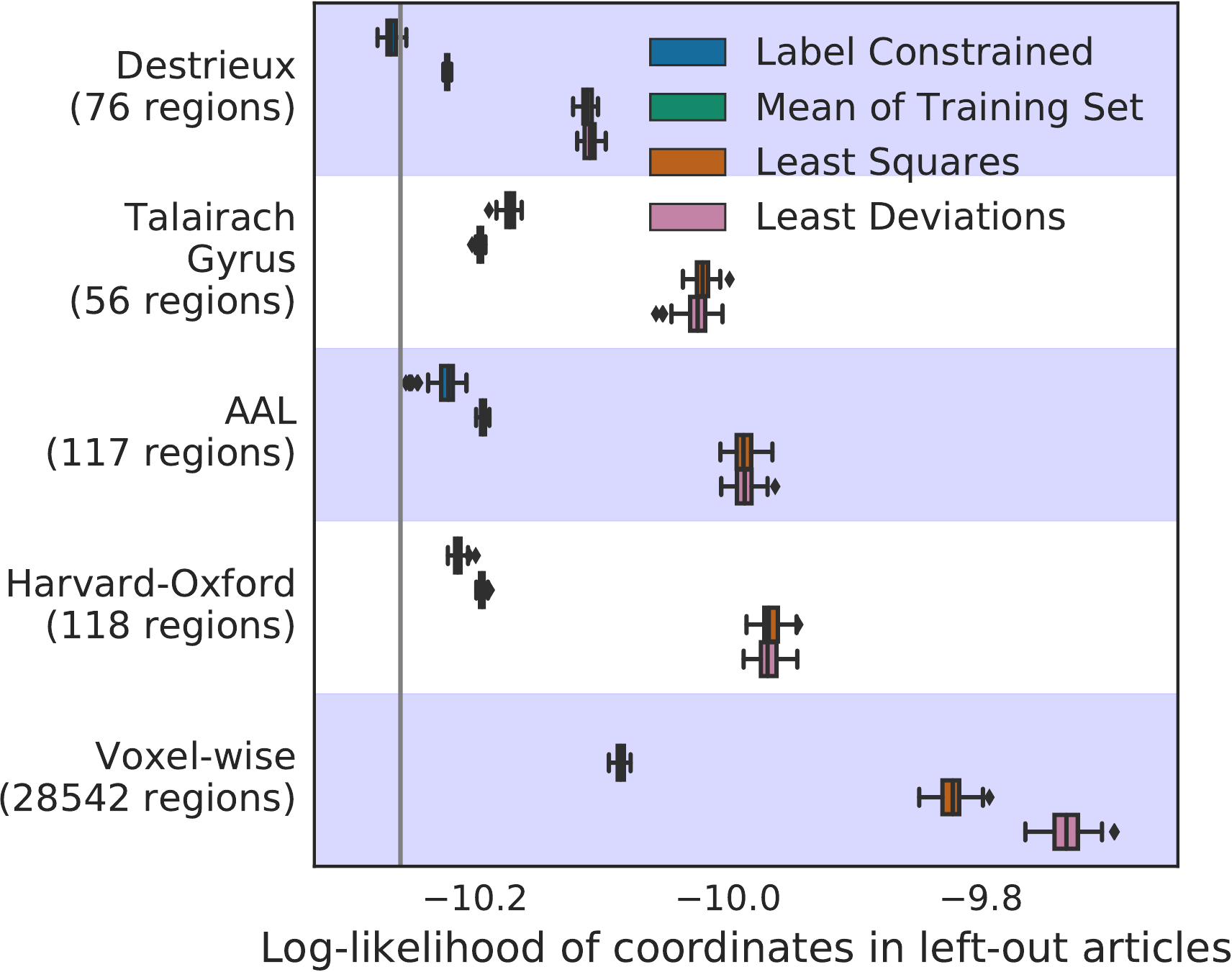}
%
%
\end{minipage}\hfill
  \begin{minipage}[c]{0.32\textwidth}
  \caption{\textbf{Log-Likelihood of coordinates reported by left-out articles in
      the predicted distribution (\cref{eq:pseudo-ll})}. The vertical line represents the
    test log-likelihood given a uniform distribution over the brain.
     Voxel-wise encoding is better than relying on any atlas. In this setting,
  $\ell_1$ regression significantly outperforms least squares.
  }
\label{fig:box-plot}
  \end{minipage}
\end{figure}

\paragraph{Comparison of atlases and models.}
We perform 100 folds of shuffle-split cross-validation (10\% in test set).
As choices of $\{\mathcal{R}_k\}$, we compare several atlases and a grid of
cubic 4-mm voxels.
%
%
%
We also compare
$\ell_1$ and $\ell_2$ regression, and label-constrained $\ell_2$.
The label-constrained encoder is
not used for the voxel grid, as it does not have labels.
As a baseline, we include a prediction based on the
average of the brain maps seen during training (i.e. independent of
the text).

\cref{fig:box-plot} gives the results:
for all models, voxel-wise encoding performs better than
any atlas. Large atlas regions regularize too much.
Despite its higher dimensionality, 
voxel-wise encoding learns better representations of anatomical terms.
The label-constrained model performs poorly, sometimes below chance, 
as the labels of a single atlas do not cover enough words and
 interactions between terms are important.
%
%
For voxel-wise encoding, $\ell_1$ regression outperforms $\ell_2$. The best
encoder is therefore learned using a  $\ell_1$ loss and a voxel partition.

\begin{figure}[b!]
  \begin{minipage}{.25\textwidth}
  \includegraphics[width=\textwidth]{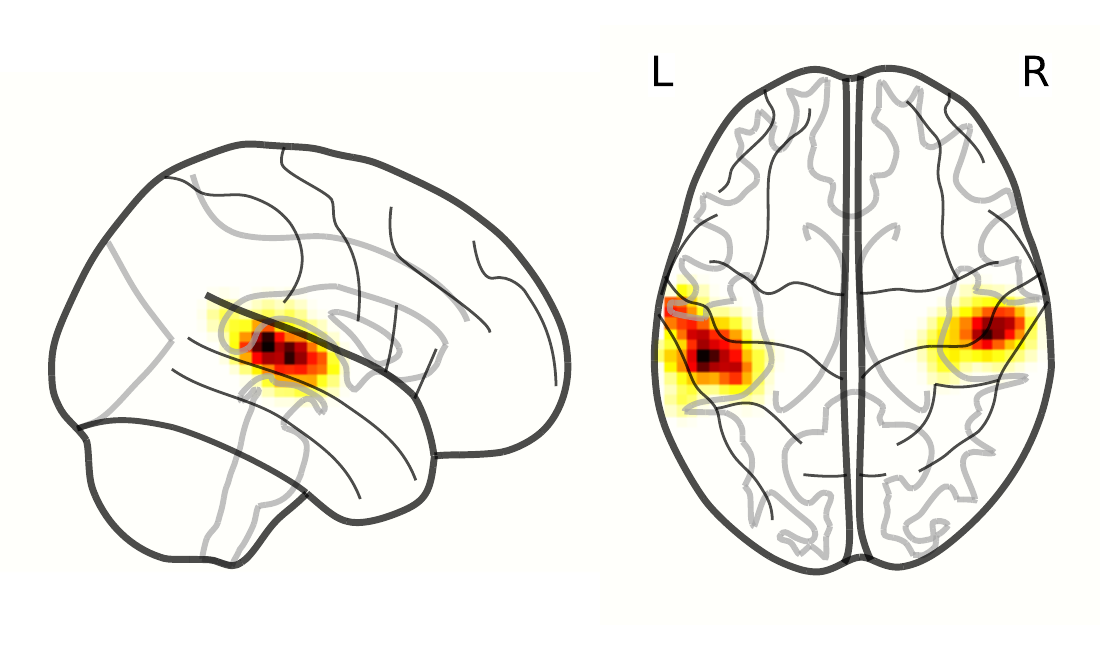}
\end{minipage}%
\begin{minipage}{.25\textwidth}
  \includegraphics[width=\textwidth]{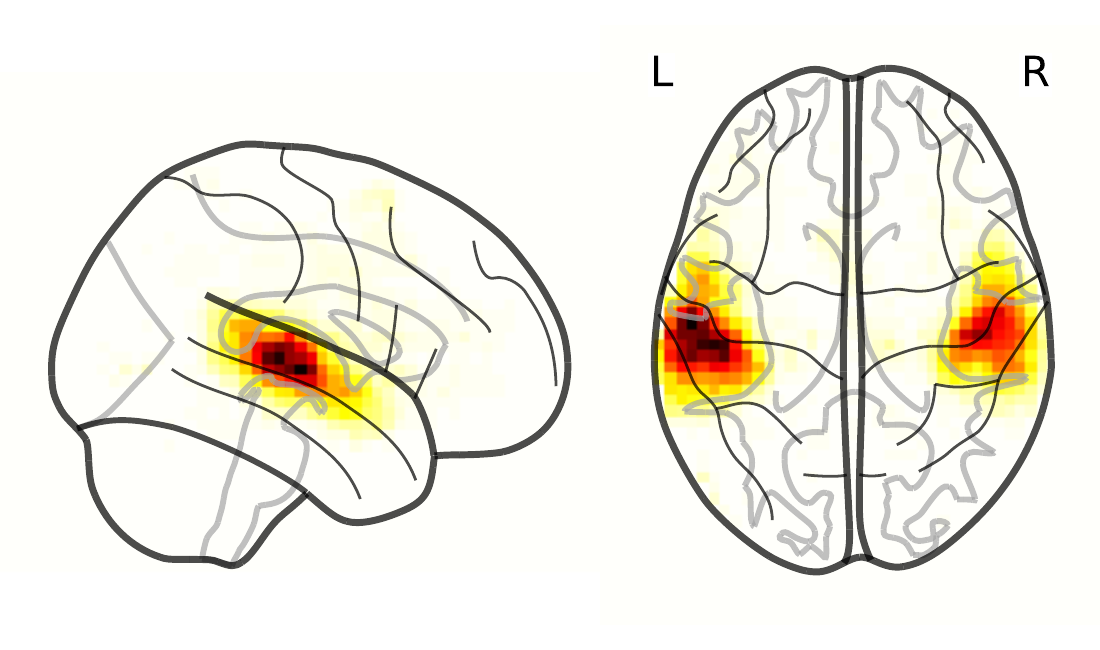}
\end{minipage}
\begin{minipage}{.5\textwidth}
  \textbf{Best prediction: }
  ``Where sound position influences sound object representations: a 7-T fMRI study''
\end{minipage}
  \begin{minipage}{.25\textwidth}
  \includegraphics[width=\textwidth]{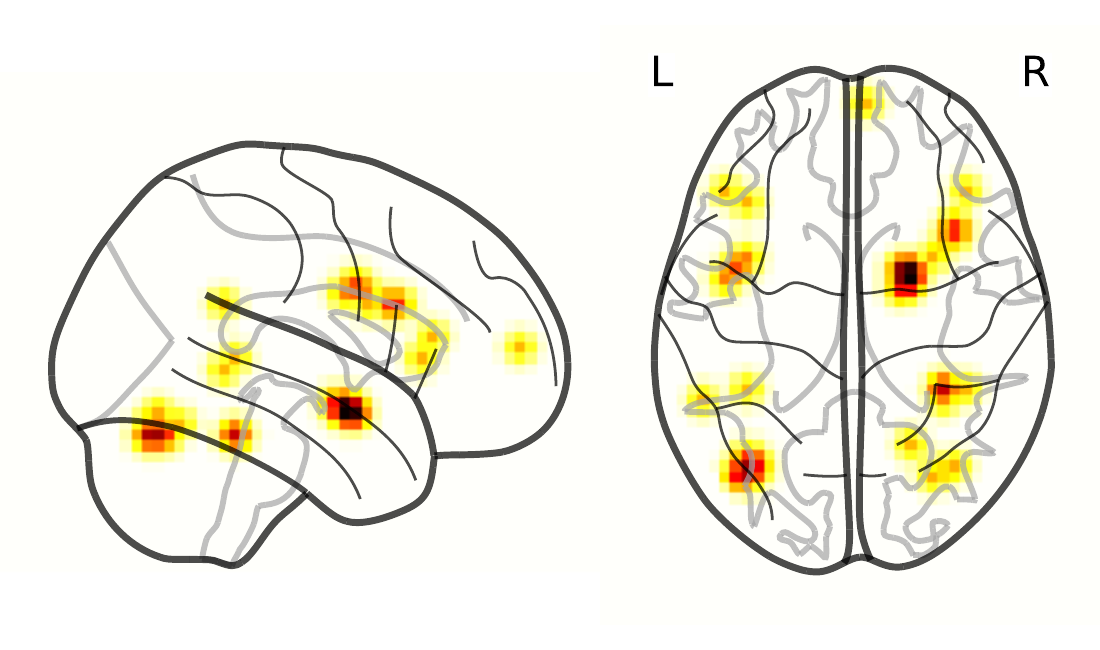}
\end{minipage}%
\begin{minipage}{.25\textwidth}
  \includegraphics[width=\textwidth]{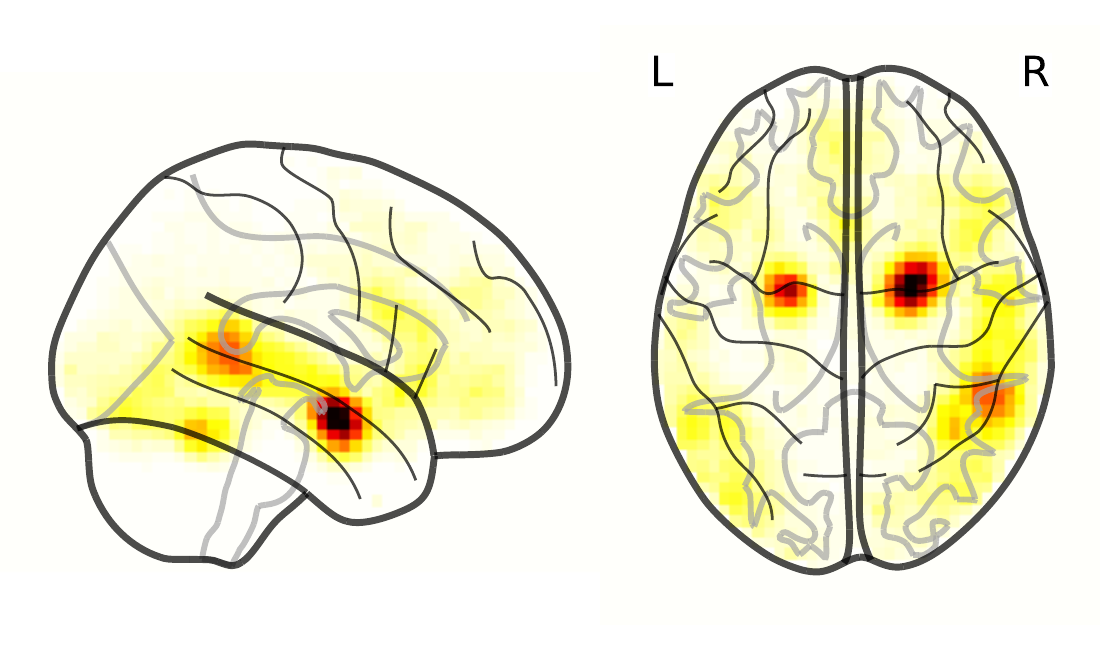}
\end{minipage}
\begin{minipage}{.5\textwidth}
  \textbf{First quartile: }
  ``Interaction of catechol O-methyltransferase and serotonin transporter genes modulates effective connectivity in a facial emotion-processing circuitry.''
\end{minipage}
\caption{True map (left) and prediction (right) for best prediction and
  $1^{st}$ quartile}
\label{fig:predictions}
\end{figure}

\paragraph{Prediction examples.} \cref{fig:predictions}
shows the true \ac{pdf} (estimated with \ac{KDE}) and the prediction for the
articles which obtained respectively the best and the first-quartile scores.
The median is shown in the supplementary material.

%

\begin{figure}[t!]
  \begin{minipage}{0.44\textwidth}
    \includegraphics[width=0.46\textwidth]{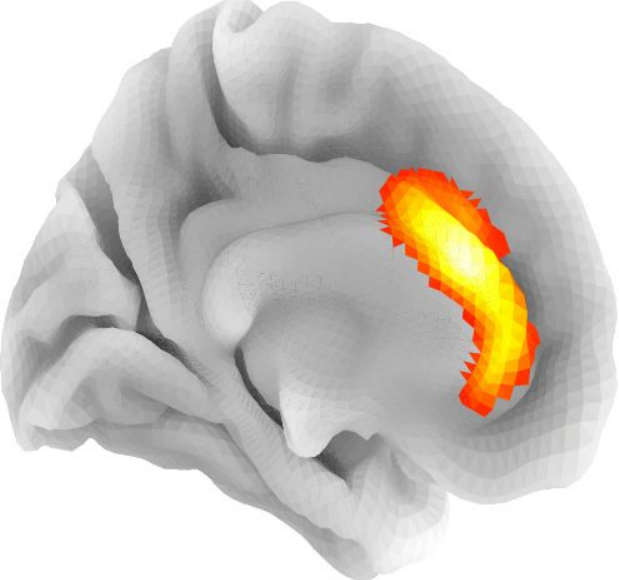}
    \includegraphics[width=0.46\textwidth]{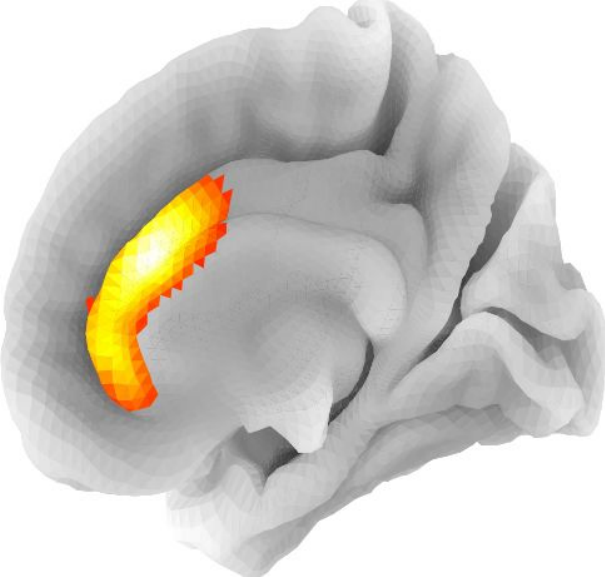}
    \caption{regression coefficient for ``anterior cingulate''}
    \label{fig:anterior-cingulate}
  \end{minipage}%
  \hfill%
  \begin{minipage}{0.5\textwidth}
  \includegraphics[width=0.29\textwidth]{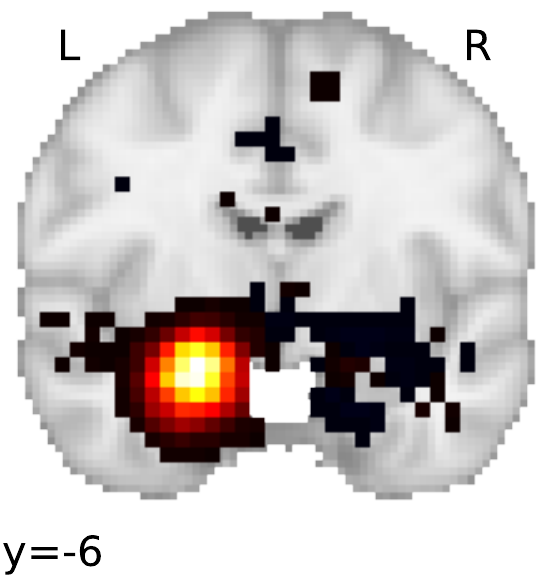}%
  \hfill%
  \includegraphics[width=0.29\textwidth]{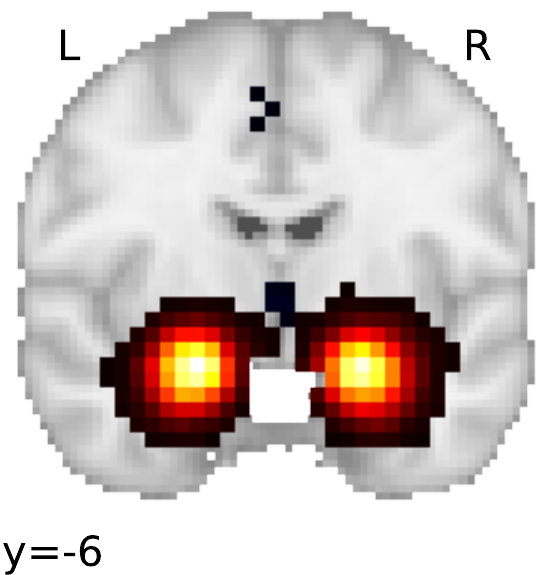}
  \hfill%
  \includegraphics[width=0.29\textwidth]{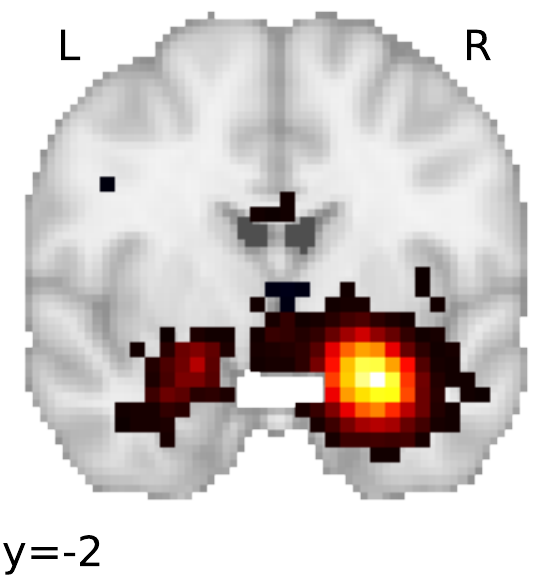}
  \caption{regression coefficients for ``left amygdala'', ``amygdala'', and ``right amygdala''%
  \label{fig:amygdala}}
  \end{minipage}
\end{figure}

\paragraph{Examples of coefficients learned by the linear regression.}
The coefficients of the linear regression (rows of $\bm{\beta}$) are the
brain maps that the model associates with each anatomical term.
For frequent terms, they are close to what experts would expect
(see for example \cref{fig:amygdala,fig:anterior-cingulate}).

\subsection{Leveraging text without coordinates: neurological examples}

\begin{figure}[b!]
  \begin{minipage}{.5\textwidth}
    \includegraphics[width=.5\textwidth]{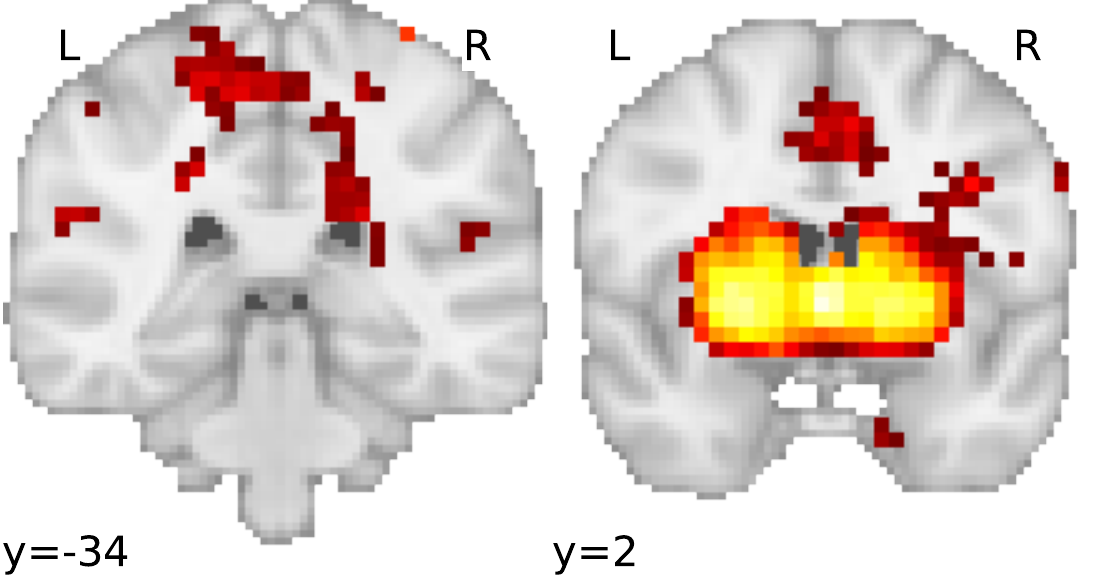}%
    \includegraphics[width=.5\textwidth]{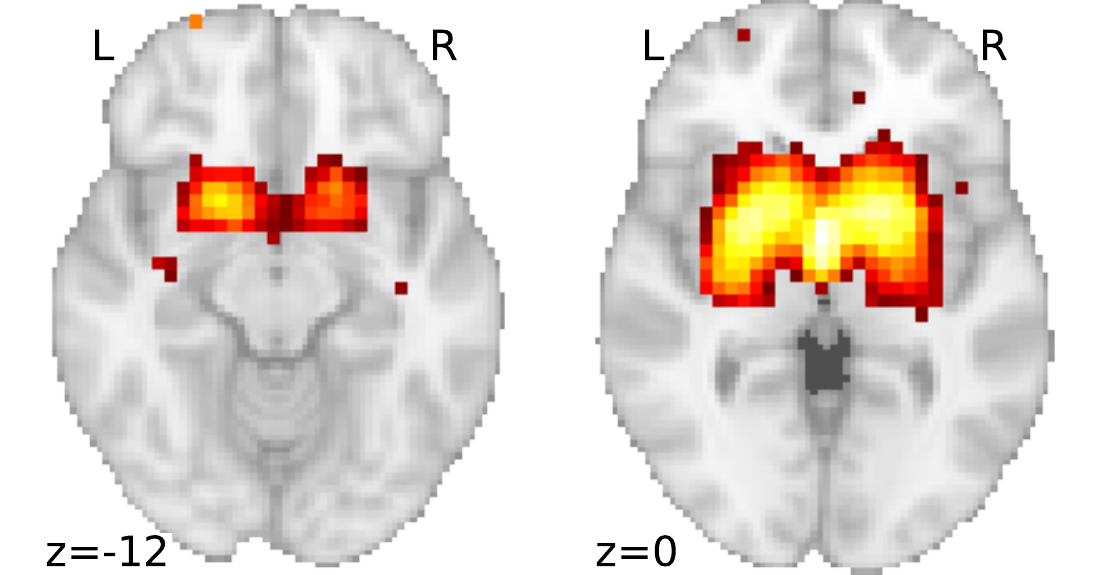}\llap{%
    \fboxsep1pt%
    \raisebox{.25\linewidth}{\colorbox{white}{\sffamily\small Huntington's
disease}}\hspace*{.2\linewidth}}%
  \llap{\rule{1pt}{.27\linewidth}}%
  \end{minipage}%
  \hspace*{2pt}%
  \begin{minipage}{.5\textwidth}
  \includegraphics[width=.5\textwidth]{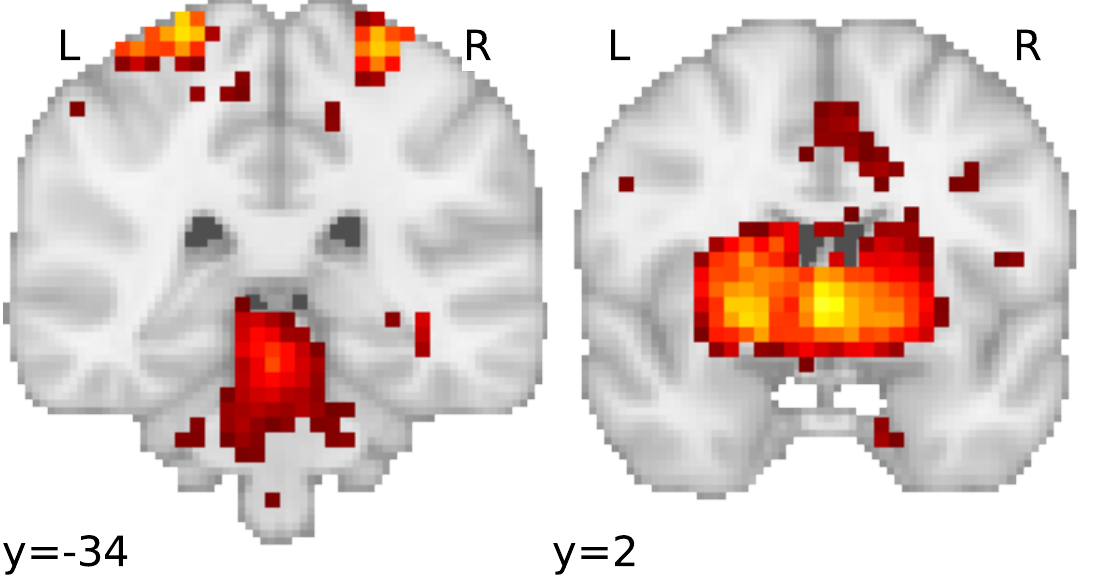}%
  \includegraphics[width=.5\textwidth]{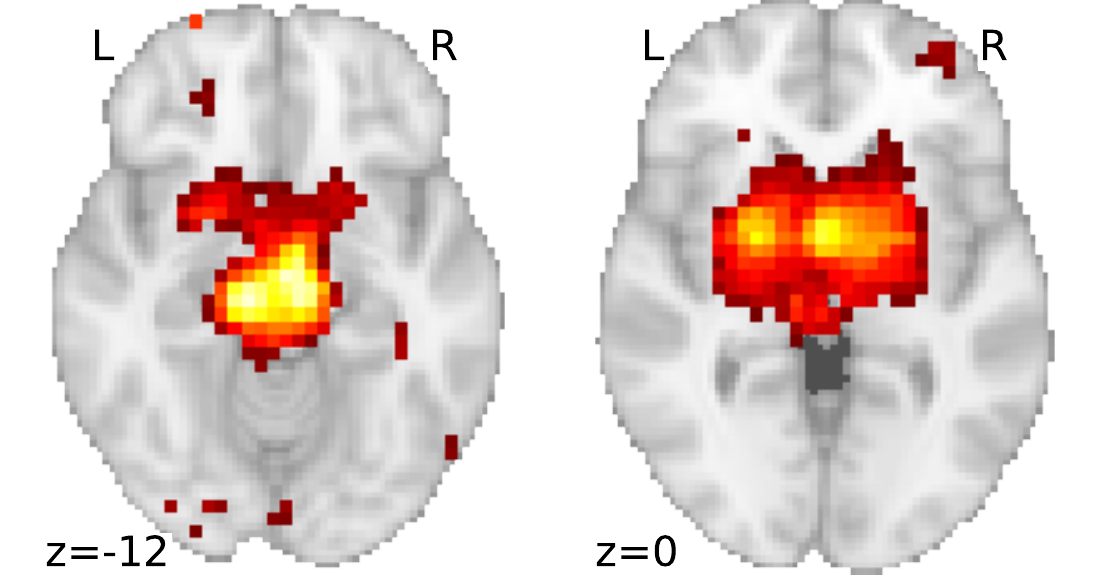}\llap{%
    \fboxsep1pt%
    \raisebox{.25\linewidth}{\colorbox{white}{\sffamily\small Parkinson's
disease}}\hspace*{.2\linewidth}}%
    \hspace*{-.1ex}%
  \end{minipage}%
  \caption{\textbf{Predicted density for Huntington's and Parkinson's.} In
    agreement with Huntington's physiopathology~\cite{walker2007huntington}, our
    method highlights the putamen, and the caudate nucleus. Also, in the case of Parkinson's~\cite{davie2008review}, the brain stem, the thalamus, and the motor
    cortex are highlighted.
  \label{fig:huntington-parkinson-slices}}
%
  \centering
  \begin{minipage}{.3\textwidth}
  \includegraphics[width=.5\textwidth]{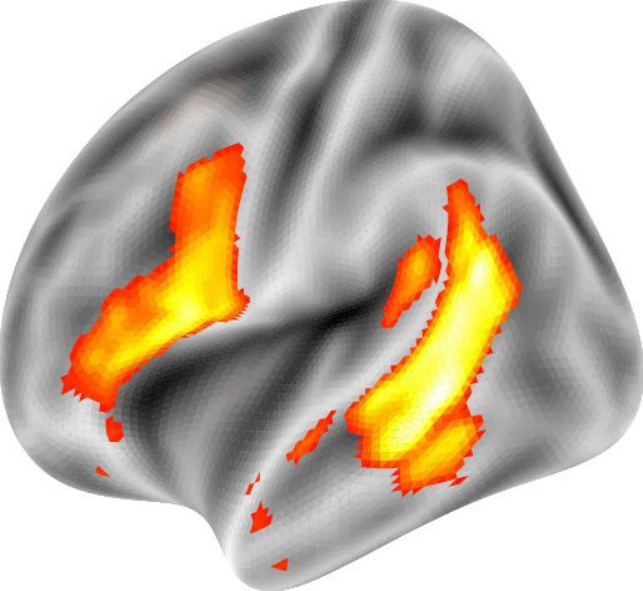}%
  \includegraphics[width=.5\textwidth]{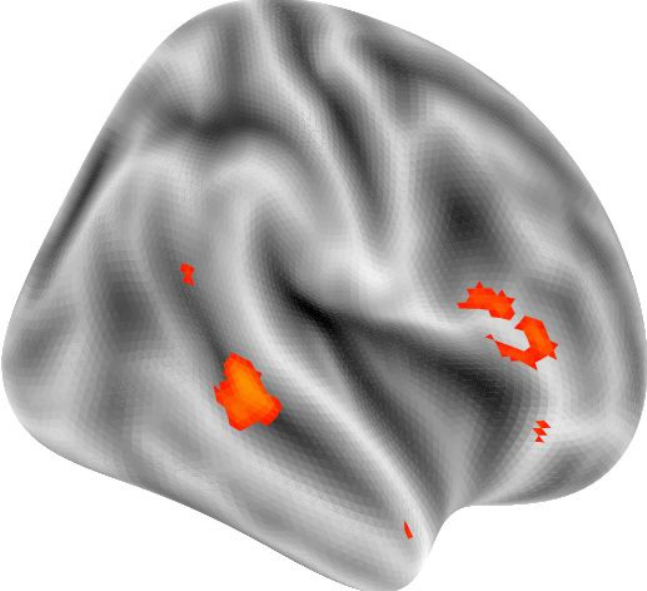}
  \end{minipage}%
  \hfill%
  \begin{minipage}{.25\textwidth}
    \includegraphics[width=\textwidth]{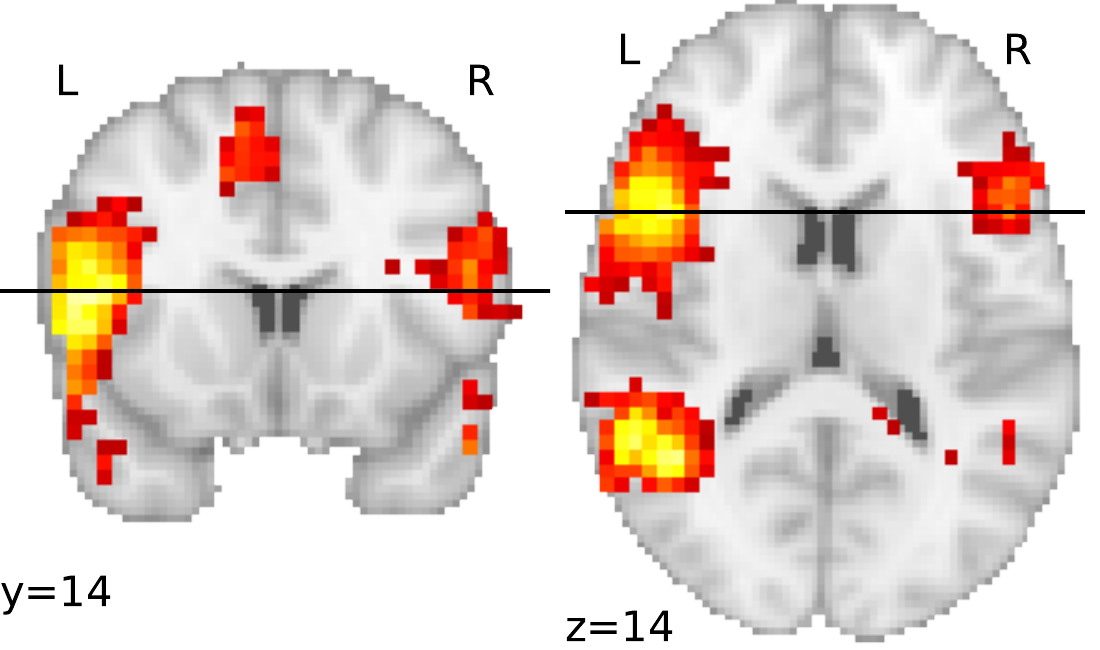}
\end{minipage}%
\hfill%
\hfill%
  \begin{minipage}{.4\textwidth}
  \caption{\textbf{Predicted density for aphasia}, centered on Broca's and
    Wernicke's areas, in agreement with the literature \cite{damasio1992aphasia}.}
  \label{fig:aphasia}
  \end{minipage}
\end{figure}

Our framework can leverage unstructured spatial
information contained in a large corpus of unannotated text.
%
%
%
To showcase this, assume that
we want to know which parts of the brain are
associated with Huntington's disease.
Our labelled corpus by itself is insufficient: only 21 documents mention the term ``huntington''.
But we use it to learn associations between anatomical terms and locations in
the brain (\cref{sec:methods}). This gives us access to
 the spatial information contained in the
unlabelled corpus, which was out of reach before (\cref{sec:encoding-performance}). We contrast the mean
encoding of articles which mention ``huntington'' against
the mean distribution (taking the difference of their $log$). Since the large
corpus
contains more information about Huntington's disease (over 400 articles mention
it), this is sufficient to see the striatum highlighted in the resulting map
(\cref{fig:huntington-parkinson-slices}, left).
\cref{fig:huntington-parkinson-slices} (right) shows the  experiment for Parkinson, and \cref{fig:aphasia} for Aphasia.

\section{Conclusion}
We have introduced a theoretical framework to translate textual description of
studies into spatial distributions over the brain.
Such a translation enables pooling together many studies which only provide text (no images or
coordinates), for statistical analysis of their results in brain space.
The statistical model gives a natural metric to validate.
This metric enables comparing representations, showing that
voxel-wise encoding is
a better approach than relying on atlases.
Building prediction models tailored to our task leads to
a linear regression with an $\ell_1$ loss (least
absolute deviation), the total-variation distance between
the true and the predicted spatial distributions.
Such a model can be trained efficiently on dozens of thousands of data
points and outperforms simpler approaches.

Applied to descriptions of pathologies that lack spatial
information, our model synthesizes accurate brain maps that
reflect the domain knowledge.
Predicting spatial distributions of medical observations
from text opens new alleys for clinical research from patient health
records and case reports.

\noindent \textbf{Acknowledgements}
This project received funding from: the European Union’s H2020 Research
Programme under Grant Agreement No. 785907 (HBP SGA2), the Metacog Digiteo
project, the MetaMRI associate team, and ERC NeuroLang.

\afterpage{\clearpage}

\bibliographystyle{splncs}
\bibliography{miccai_biblio.bib}

\clearpage

\appendix

\textbf{\Large{Supplementary material}}

\section{Factorization of the loss function}\label{sec:factorize-loss}

\begin{IEEEeqnarray}{rCl}
   \int_{\mathbb{R}^3} \delta(\hat{p}(z) - q(z))dz & = &%
  \int_{\mathbb{R}^3} \delta(\hat{p}(z) - q(z))\sum_{k=0}^m\mathbb{I}_k(z)dz \\
  & = & \int_{\mathbb{R}^3} \delta(\hat{p}(z) - q(z))\sum_{k=1}^m\mathbb{I}_k(z)dz ~~
  \footnote{because p and q are null in the background and  $\delta(0, 0) = 0$} \\
  & = & \sum_{k=1}^m\int_{\mathbb{R}^3} \delta(\hat{\bm{y}}_k r_k(z) -
  \sum_{t=1}^d\bm{x}_{t}\bm{\beta}_{t,k}r_k(z))\mathbb{I}_k(z) dz  ~~
  \footnote{because $\mathbb{I}_k \neq 0 \implies r_{k'} = 0 ~~ \forall k' \neq k$}\\
   & = & \sum_{k=1}^m\int_{\mathbb{R}^3} \delta(\frac{\hat{\bm{y}}_k}{v_k} - \frac{\sum_{t=1}^d\bm{x}_{t}\bm{\beta}_{t,k}}{v_k})\mathbb{I}_k(z) dz \\
  & = &
    \sum_{k=1}^m v_k \delta(\frac{\hat{\bm{y}}_k}{v_k} - \frac{\sum_{t=1}^d\bm{x}_{t}\bm{\beta}_{t,k}}{v_k})
\end{IEEEeqnarray}

\section{Derivation of the dual of penalized least-deviations}\label{sec:l1-loss-dual}

We have the minimization problem:
\begin{equation}
  \hat{\bm{\beta}} = \argmin_{\bm{\beta}}\left( \|\hat{\bm{Y}} - \bm{X}\bm{\beta}\|_1 + \lambda \|\bm{\beta}\|_2^2  \right)
\end{equation}
where $\bm{X} = (\bm{x}_i) \in \mathbb{R}^{n\times d}$ and $\hat{\bm{Y}} = (\hat{\bm{y}}_i) \in \mathbb{R}^{n \times m}$
  and  $\lambda \in \mathbb{R}_+$.

\noindent The problem is equivalent to:
\begin{align}
  \argmin_{\bm{Z}, \bm{\beta}}\left( \|\bm{Z}\|_1 + \lambda \|\bm{\beta}\|_2^2 \right)   \\
   \text{ s.t. } \hat{\bm{Y}} - \bm{X}\bm{\beta} - \bm{Z} = 0
\end{align}
Introducing the dual variable $\bm{\nu} \in \mathbb{R}^{n\times m}$, the Lagrangian is:
\begin{equation}
  L(\bm{Z}, \bm{\beta}, \bm{\nu}) = \|\bm{Z}\|_1 + \lambda \|\bm{\beta}\|_2^2 + Tr(\bm{\nu}^T(\hat{\bm{Y}} - \bm{X}\bm{\beta} - \bm{Z}))
\end{equation}
The derivative with respect to $\bm{\beta}$ is
\begin{equation}
  2 \lambda \bm{\beta} - \bm{X}^T\bm{\nu}
\end{equation}
So minimizing with respect to $\bm{\beta}$ yields $\bm{\beta} = \frac{\bm{X}^T\bm{\nu}}{2\lambda}$
and
\begin{equation}
  \min_{\bm{\beta}}L(\bm{Z}, \bm{\beta}, \bm{\nu}) = \|\bm{Z}\|_1 + \Tr(\bm{\nu}^T\hat{\bm{Y}} - \bm{\nu}^T\bm{Z} - \frac{1}{4\lambda} \bm{\nu}^TXX^T\bm{\nu})
\end{equation}
The dual norm of the $l_1$ norm is $l_{\infty}$, so minimizing with respect to
$\bm{Z}$ we get the Lagrange dual function
\begin{IEEEeqnarray}{rCl}
  g(\bm{\nu}) = \min_{\bm{Z}, \bm{\beta}}L(\bm{Z}, \bm{\beta}, \bm{\nu}) & = \begin{cases}
    Tr(\bm{\nu}^T\hat{\bm{Y}} - \frac{1}{4\lambda}\bm{\nu}^TXX^T\bm{\nu}) & \text{ if } \|\bm{\nu}\|_{\infty} \leq 1 \\
    - \infty & \text{otherwise} \end{cases}
\end{IEEEeqnarray}
%
%
%
The dual problem is:
\begin{align}
  \hat{\bm{\nu}} = \argmax_{\bm{\nu}}\left( Tr(\bm{\nu}^T\hat{\bm{Y}} -
\frac{1}{4\lambda}\bm{\nu}^TXX^T\bm{\nu}) \right)   \quad
   \text{ s.t. } \|\bm{\nu}\|_\infty \leq 1
\end{align}
$g$ is differentiable; its gradient is
\begin{equation}
  \nabla_g(\bm{\nu}) = \hat{\bm{Y}} - \frac{1}{2\lambda} \bm{X}\bm{X}^T \bm{\nu}
\end{equation}
And we solve this problem using an efficient algorithm: L-BFGS.
Then we get back the primal solution as $\hat{\bm{\beta}} = \frac{\bm{X}^T\hat{\bm{\nu}}}{2\lambda}$.
In practice, since data must be centered and normalized, the mean and
scale of $\bm{X}$ appear in these formulas so that we do not break sparsity of $\bm{X}$:
$g$ is written
\begin{equation}
Tr(\bm{\nu}^T\tilde{\bm{Y}} - \frac{1}{4\lambda}\bm{K}^T\bm{K})
\end{equation}
With
\begin{equation}
  \bm{K} = (\tilde{\bm{X}} - \bar{\bm{x}})^T\bm{\nu} = \tilde{\bm{X}}^T\bm{\nu} - \bar{\bm{x}} \odot \sum_{i=1}^n\bm{\nu}_i
\end{equation}
Where $\tilde{\bm{X}} \in \mathbb{R}^{n \times d}$ is $\bm{X}$ divided by $n$ times the
variance of its columns, $\bar{\bm{x}} \in \mathbb{R}^d$ is the mean of the
columns of $\bm{X}$ divided by the same quantity, $\tilde{\bm{Y}}$ is the centered and
normalized $\hat{\bm{Y}}$, and $\odot$ is the Hadamard
product. This is fast to compute because $\tilde{\bm{X}}$ is sparse (in practice,
over 97\% of entries are null). In a similar way, the gradient becomes
\begin{equation}
  -\tilde{\bm{Y}} + \frac{1}{2\lambda}(\tilde{\bm{X}}\bm{K} - \bar{\bm{x}}\bm{K})
\end{equation}
and $\bm{\beta}$ is given by
\begin{equation}
  \tilde{\bm{X}}^T \bm{\nu} - \bar{\bm{x}} \odot \sum_{i=1}^n \bm{\nu}_i
\end{equation}

\newpage 

\section{More extensive atlas comparison}


\begin{figure}
  \includegraphics[width=1.\textwidth]{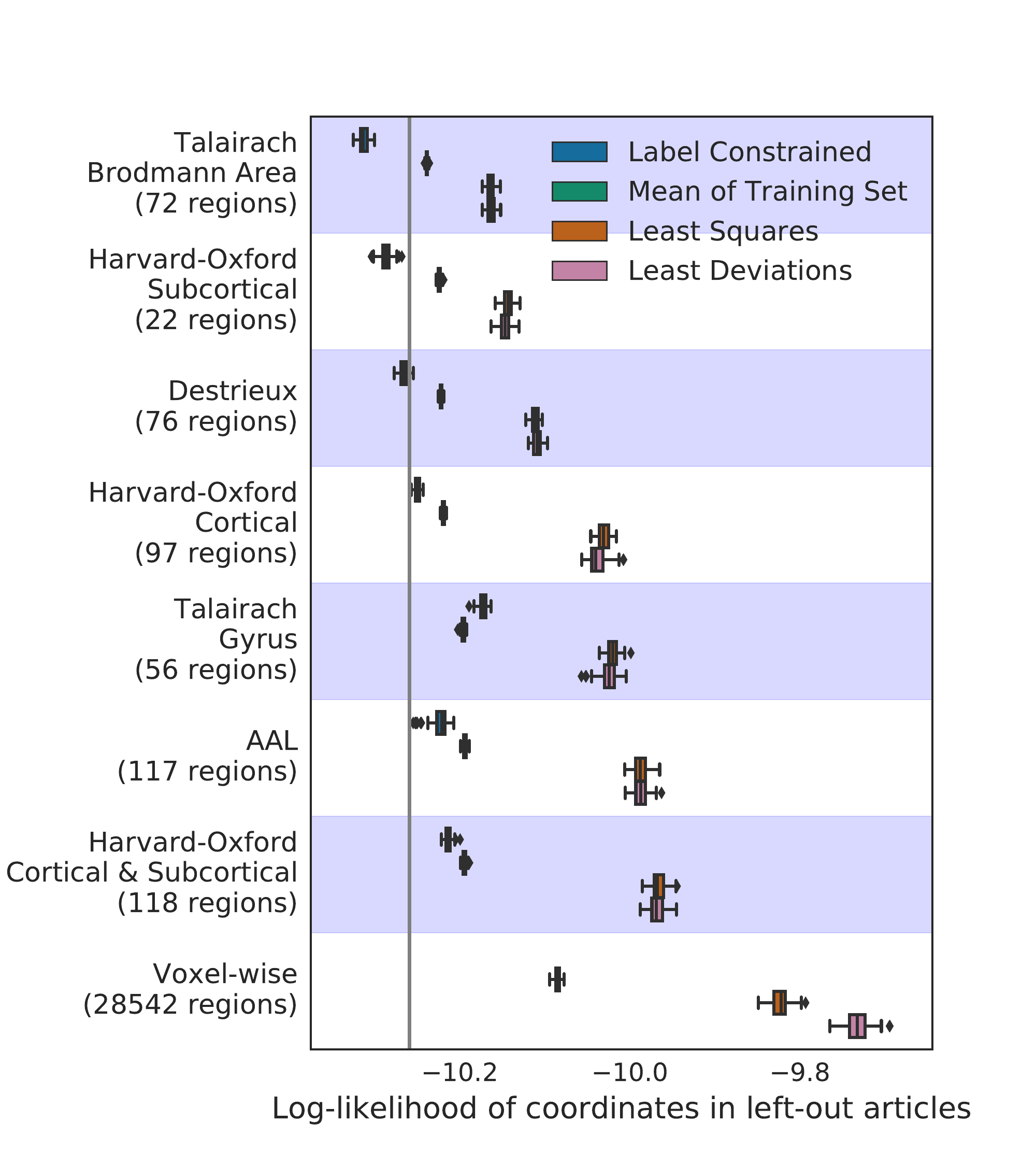}
  \caption{Log-Likelihood of coordinates reported by left-out articles in
      the predicted distribution.}
\label{fig:box-plot-2}
\end{figure}

\clearpage

\section{Example predictions}\label{sec:appendix-example-predictions}

\subsection{Best prediction}
\noindent\textbf{Title:}``Where sound position influences sound object representations: a
7-T fMRI study'' \citeannex{van2011sound}

\noindent\textbf{Abstract:}
``Evidence from human and non-human primate studies supports a dual-pathway model
of audition, with partially segregated cortical networks for sound recognition
and sound localisation, referred to as the What and Where processing streams. In
normal subjects, these two networks overlap partially on the supra-temporal
plane, suggesting that some early-stage auditory areas are involved in
processing of either auditory feature alone or of both. Using high-resolution
7-T fMRI we have investigated the influence of positional information on sound
object representations by comparing activation patterns to environmental sounds
lateralised to the right or left ear. While unilaterally presented sounds
induced bilateral activation, small clusters in specific non-primary auditory
areas were significantly more activated by contra-laterally presented stimuli.
Comparison of these data with histologically identified non-primary auditory
areas suggests that the coding of sound objects within early-stage auditory
areas lateral and posterior to primary auditory cortex AI is modulated by the
position of the sound, while that within anterior areas is not.''
\begin{figure}
  \begin{minipage}{.25\textwidth}
  \includegraphics[width=\textwidth]{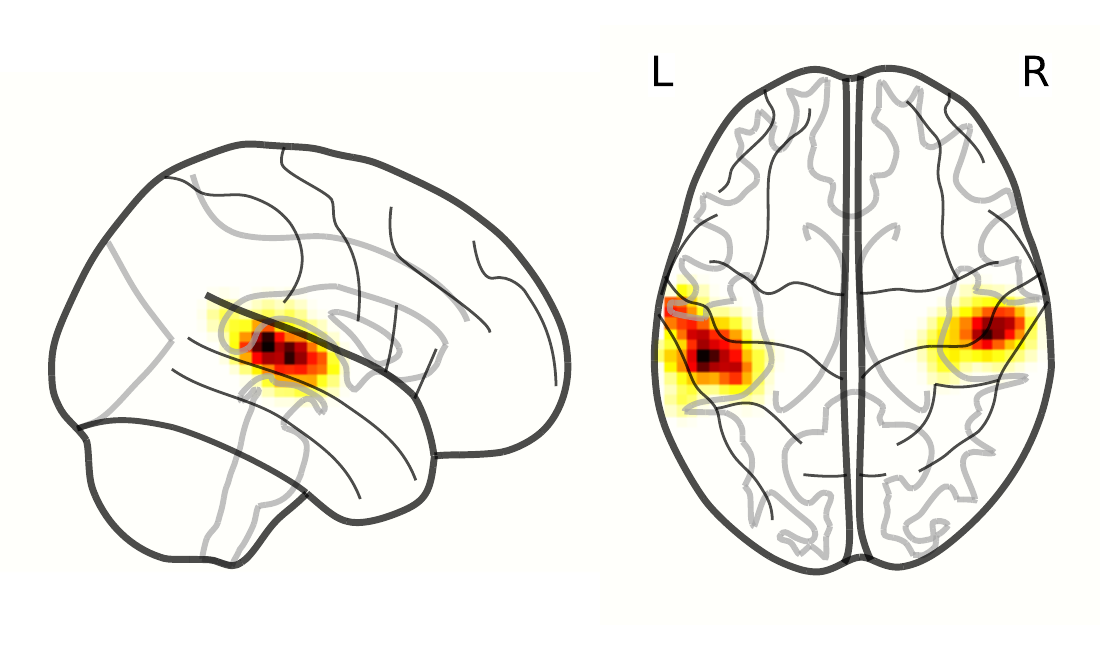}
\end{minipage}%
\begin{minipage}{.25\textwidth}
  \includegraphics[width=\textwidth]{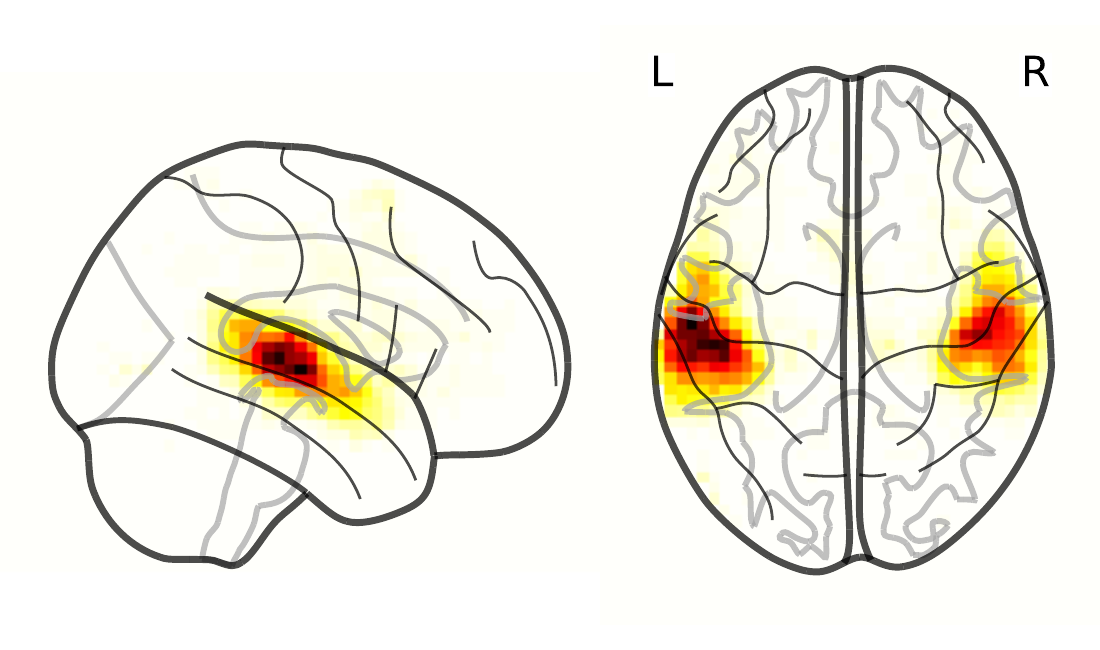}
\end{minipage}
\begin{minipage}{.5\textwidth}
  True map (left) and prediction (right)
\end{minipage}
 \end{figure}
\vspace{-8mm} 
\subsection{First quantile}

\noindent\textbf{Title:}``Interaction of catechol O-methyltransferase and serotonin transporter genes modulates effective connectivity in a facial emotion-processing circuitry.''\citeannex{surguladze2012interaction}

\noindent\textbf{Abstract:}
``Imaging genetic studies showed exaggerated blood oxygenation level-dependent
response in limbic structures in carriers of low activity alleles of serotonin
transporter-linked promoter region (5-HTTLPR) as well as catechol
O-methyltransferase (COMT) genes. This was suggested to underlie the
vulnerability to mood disorders. To better understand the mechanisms of
vulnerability, it is important to investigate the genetic modulation of
frontal-limbic connectivity that underlies emotional regulation and control. In
this study, we have examined the interaction of 5-HTTLPR and COMT genetic
markers on effective connectivity within neural circuitry for emotional facial
expressions. A total of 91 healthy Caucasian adults underwent functional
magnetic resonance imaging experiments with a task presenting dynamic emotional
facial expressions of fear, sadness, happiness and anger. The effective
connectivity within the facial processing circuitry was assessed with Granger
causality method. We have demonstrated that in fear processing condition, an
interaction between 5-HTTLPR (S) and COMT (met) low activity alleles was
associated with reduced reciprocal connectivity within the circuitry including
bilateral fusiform/inferior occipital regions, right superior temporal
gyrus/superior temporal sulcus, bilateral inferior/middle prefrontal cortex and
right amygdala. We suggest that the epistatic effect of reduced effective
connectivity may underlie an inefficient emotion regulation that places these
individuals at greater risk for depressive disorders.''
\begin{figure}
  \begin{minipage}{.25\textwidth}
  \includegraphics[width=\textwidth]{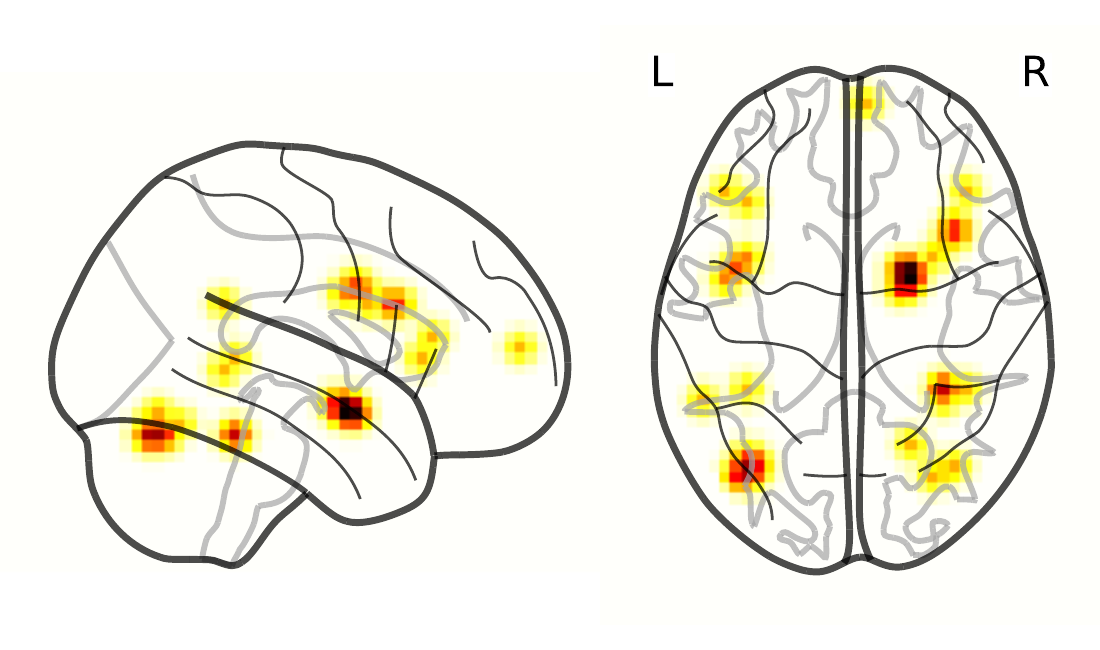}
\end{minipage}%
\begin{minipage}{.25\textwidth}
  \includegraphics[width=\textwidth]{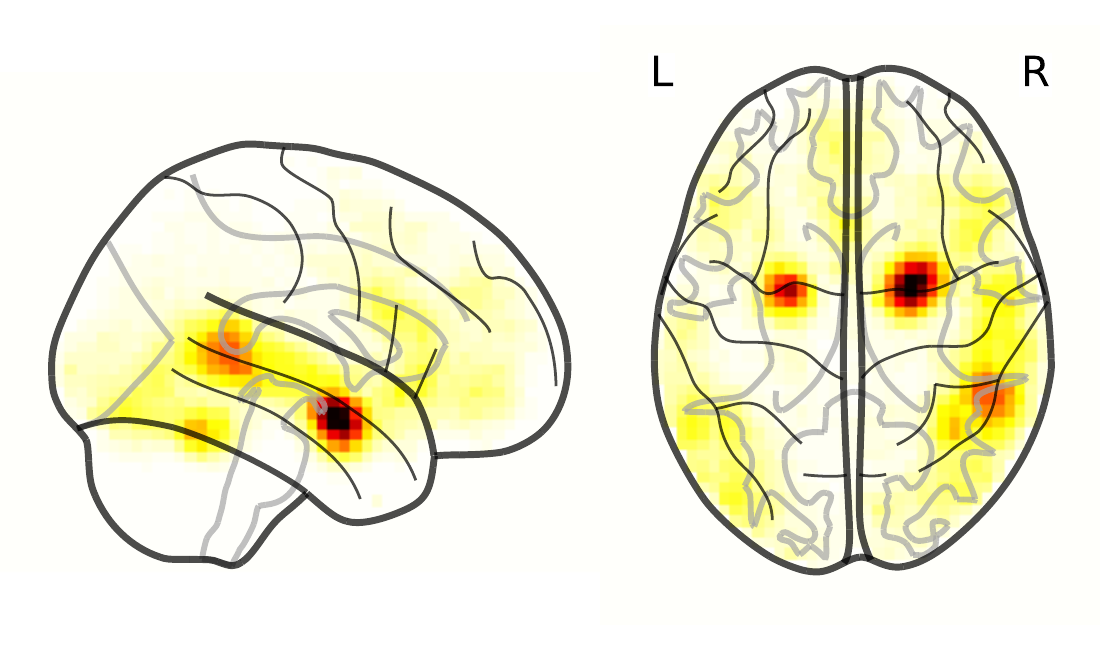}
\end{minipage}
\begin{minipage}{.5\textwidth}
  True map (left) and prediction (right)
\end{minipage}
\end{figure}
\vspace{-8mm} 
\subsection{Median}

\noindent\textbf{Title:}
``How specifically are action verbs represented in the neural motor system: an fMRI study.''\citeannex{van2010specifically}

\noindent\textbf{Abstract:}
``Embodied accounts of language processing suggest that sensorimotor areas,
generally dedicated to perception and action, are also involved in the
processing and representation of word meaning. Support for such accounts comes
from studies showing that language about actions selectively modulates the
execution of congruent and incongruent motor responses (e.g., Glenberg \&
Kaschak, 2002), and from functional neuroimaging studies showing that
understanding action-related language recruits sensorimotor brain areas (e.g.
Hauk, Johnsrude, \& Pulvermueller, 2004). In the current experiment we explored
the basis of the neural motor system's involvement in representing words
denoting actions. Specifically, we investigated whether the motor system's
involvement is modulated by the specificity of the kinematics associated with a
word. Previous research in the visual domain indicates that words denoting basic
level category members lacking a specific form (e.g., bird) are less richly
encoded within visual areas than words denoting subordinate level members (e.g.,
pelican), for which the visual form is better specified (Gauthier, Anderson,
Tarr, Skudlarski, \& Gore, 1997). In the present study we extend these findings
to the motor domain. Modulation of the BOLD response elicited by verbs denoting
a general motor program (e.g., to clean) was compared to modulation elicited by
verbs denoting a more specific motor program (e.g., to wipe). Conform with our
hypothesis, a region within the bilateral inferior parietal lobule, typically
serving the representation of action plans and goals, was sensitive to the
specificity of motor programs associated with the action verbs. These findings
contribute to the growing body of research on embodied language representations
by showing that the concreteness of an action-semantic feature is reflected in
the neural response to action verbs.''

\begin{figure}
  \begin{minipage}{.25\textwidth}
  \includegraphics[width=\textwidth]{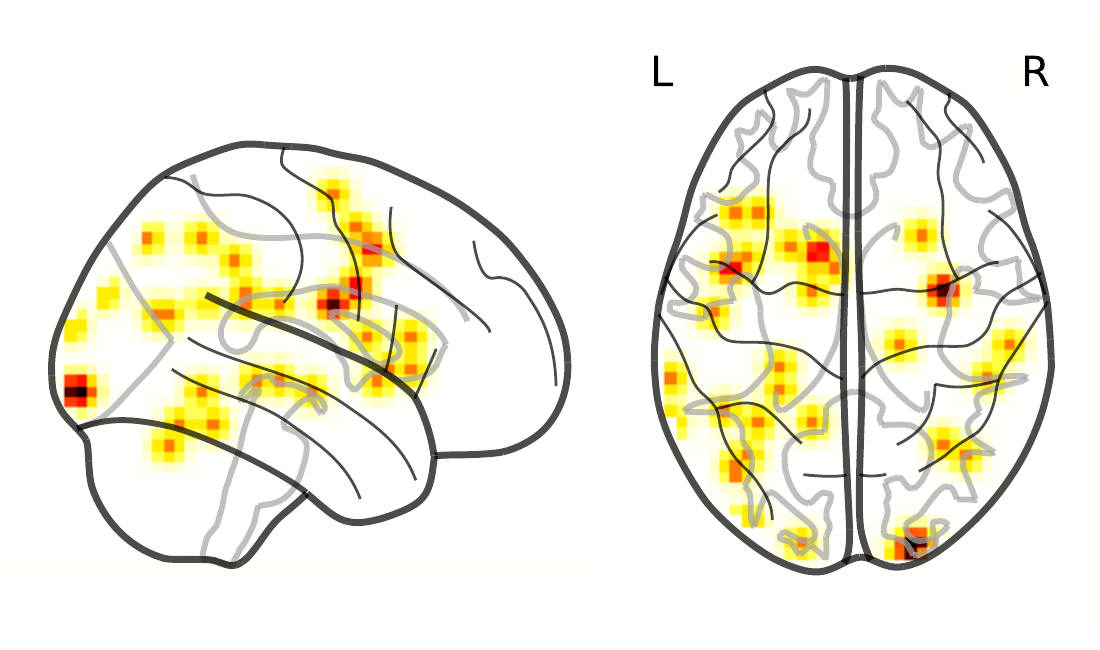}
\end{minipage}%
\begin{minipage}{.25\textwidth}
  \includegraphics[width=\textwidth]{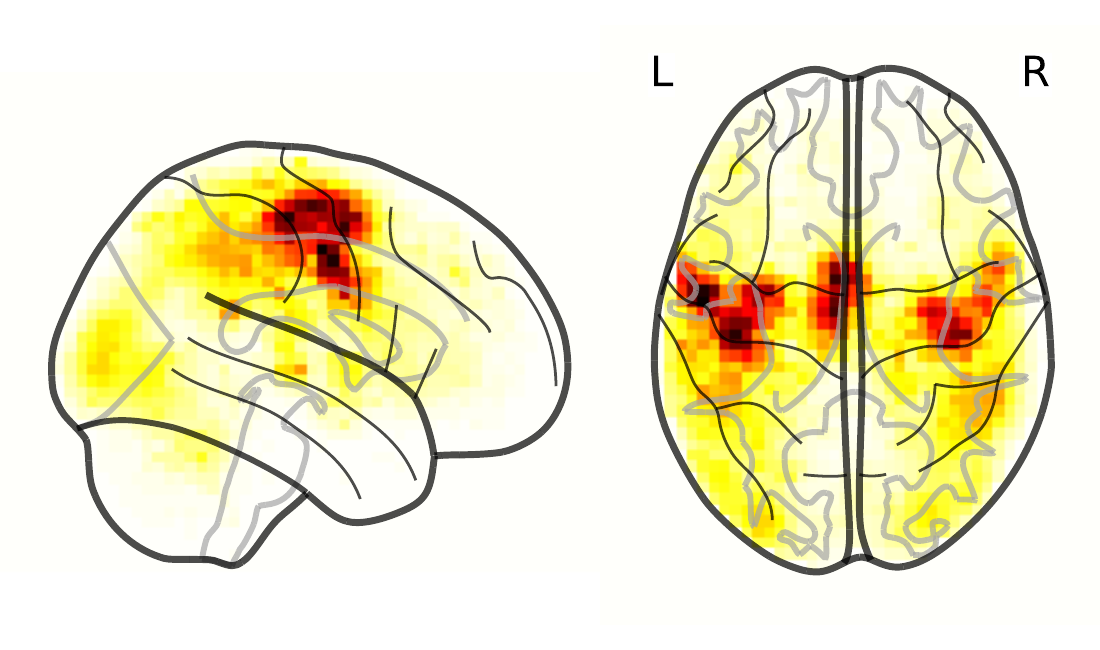}
\end{minipage}
\begin{minipage}{.5\textwidth}
  True map (left) and prediction (right)
\end{minipage}
\end{figure}

\section{Fast \acl{KDE} with convolutions}\label{sec:kde-implementation}

\acl{KDE} is a non-parametric way to estimate the \ac{pdf} of a random
variable, given a sample (possibly weighted). In our case the sample
is the list of coordinates provided by a study. We use \ac{KDE} to draw the brain
map associated with each study. A reference on density estimation can
be found in \citeannex{silverman1986density} or \citeannex{simonoff2012smoothing}.
Once we have chosen a kernel $\phi$ (a function that sums to 1, symmetric and
non-negative), we define the \emph{rescaled kernel}:

\begin{equation}\label{eq:rescaled-kernel-nd}
  \phi_{\bm{H}}(\bm{u}) = |\bm{H}|^{-\frac{1}{2}}\phi\left(\bm{H}^{-\frac{1}{2}}\bm{u}\right)
\end{equation}
where $|\cdot|$ is the determinant, and the smoothing parameter
$\bm{H} \in \mathbb{R}^{d \times d}$ is a $(d \times d)$ symmetric positive
definite matrix, called the \emph{bandwidth matrix}.

The estimate of the \ac{pdf} $\hat{p}$ is then given by:

\begin{equation}\label{eq:kde-nd}
  \hat{p}(\bm{v})  =  \frac{1}{c}\sum_{a=1}^c \omega_a\phi_{\bm{H}}\left(\bm{v} - \bm{l}_a\right)
\end{equation}
where $\{\bm{l}_a, a = 1\dots c\}$ is the sample and
$\{\omega_a, a = 1 \dots c\}$ are the weights associated with the
${\bm{l}_a}$ and sum to $c$.

Note that if $\phi$ is the standard multivariate density (which is the
kernel we use in practice), $\phi_{\bm{H}}$ is the multivariate density of
mean $0$ and covariance $\bm{H}$:
\begin{equation}
  \phi(\bm{u}) = \frac{1}{\sqrt{(2\pi)^d}}\exp\left(-\frac{1}{2} \bm{u}^T\bm{u}\right)
\end{equation}
and
\begin{IEEEeqnarray}{rCl}
  \phi_{\bm{H}}(\bm{u}) & = & |\bm{H}|^{-\frac{1}{2}}\phi\left(\bm{H}^{-\frac{1}{2}}\bm{u}\right) \\
  & = & \frac{1}{\sqrt{(2\pi)^d|\bm{H}|}}\exp\left(-\frac{1}{2} \bm{u}^T(\bm{H}^{-\frac{1}{2}})^T\bm{H}^{-\frac{1}{2}}\bm{u}\right) \\
  & = & \frac{1}{\sqrt{(2\pi)^d|\bm{H}|}}\exp\left(-\frac{1}{2} \bm{u}^T\bm{H}^{-1}\bm{u}\right)
\end{IEEEeqnarray}


In practice, we use a multivariate Gaussian kernel and isotropic
smoothing (which is usually the case in neuroimaging):
\begin{equation}
  \bm{H} = \begin{bmatrix}h^2&&\\&h^2&\\&&h^2\end{bmatrix}
\end{equation}
So \Cref{eq:kde-nd} can be rewritten as
\begin{equation}
    \hat{p}(\bm{v}) = \frac{1}{ch^d}\sum_{a=1}^c \omega_a\phi\left(\frac{\bm{v} - \bm{l}_a}{h}\right)
\end{equation}
with $d = 3$ is the dimension of the vector space in which our samples live.

A naive computation of this sum would be expensive.
But since the estimate is the convolution of the
kernel with the data, it can be computed efficiently with Fourier
transforms \citeannex{wand1994fast,silverman1986density}. In the general
\ac{KDE} setting, the sample points lie in a continuous space which
needs to be binned in order to compute the discrete convolution. The
grid we choose for binning will define the voxels of the brain map we
compute. Two strategies are popular for computing the binned
approximation of a weighted sample: simple binning and linear
binning. Simple binning consists in assigning the whole weight of each
sample point to the closest grid node. Linear binning distributes the
weight of the sample point over the neighbouring nodes, e.g., in the
one-dimensional case, if $\bm{v}$ lies between $i\delta$ and $(i+1)\delta$,
($\delta$ being the grid increment) the fraction of $\bm{v}$'s weight that
$i\delta$ will receive is $\frac{(i+1)\delta - \bm{v}}{\delta}$. We simply
assign the weights of the activation peaks to the voxel in which they
lie. In other words, if an activation peak lies between $i\delta$ and
$(i+1)\delta$, its whole weight will go to $i\delta$. Thus we compute
the weights $w_{i,j,k}$ associated with the nodes of our grid (the
voxels of our image) as
\begin{equation}
  w_{i,j,k} = \sum_{a \in \mathcal{N}(i,j,k)}\omega_{a}
\end{equation}
where $\omega_{a}$ is the weight associated with sample point $l_a$ and where
$a$ belongs to the neighbourhood $\mathcal{N}(i,j,k)$ if and only if, writing
$l_a = (x, y, z)$ the coordinates in $\mathbb{R}^3$ of $l_a$,
\begin{align*}
  i \delta_x \leq x < (i+1) \delta_x \\
  j \delta_y \leq y < (j+1) \delta_y \\
  k \delta_z \leq z < (k+1) \delta_z
\end{align*}
where $\delta_x$, $\delta_y$ and $\delta_z$ are the side lengths of
the voxels.

The ${w_{i,j,k}}$ are sometimes called the \emph{bin counts}.
\Cref{eq:kde-nd} can be rewritten in voxel space as:
\begin{equation}
  \hat{p}((i', j', k')) = \frac{1}{c}\sum_{i=1}^{n_x}\sum_{j=1}^{n_y}\sum_{k=1}^{n_z} w_{i,j,k}\phi_{\bm{H}}\left( |i' - i|\delta_x, |j' - j|\delta_y, |k' - k|\delta_z \right)
\end{equation}
where $n_x$, $n_y$ and $n_z$ are the dimensions of the
image.
This is the convolution of two images, one, $\bm{K}$, containing kernel
evaluations at multiples of the voxel dimensions, and one, $\bm{W}$
containing the bin counts.

In \citeannex{wand1994fast}, the authors present a scheme for zero-padding
the matrices to convolve in a way that ensures very fast computation
of their Fourier Transforms in the one-dimensional and two-dimensional
cases; it generalizes immediatly to higher dimensions and we have used
it for our three-dimensional images. This scheme is only valid when
the bandwidth matrix $\bm{H}$ is diagonal; \citeannex{gramacki2016fft}
recently described another padding which is adapted for unconstrained
bandwidth matrices. However, the smoothing matrix we use is indeed
diagonal and the procedure described in \citeannex{wand1994fast} can be applied in
our case. We use the notation $\diag(\bm{H}) = (h_x, h_y, h_z)$

We use a Gaussian kernel, which has infinite support. However, it
decreases very rapidly and we can approximate it by the truncated
kernel, taking $\phi_{\bm{H}}(i\delta_x, j\delta_y, k\delta_z)$ to be $0$
when $|\frac{i\delta_x}{h_x}| \geq 5$,
$|\frac{j\delta_y}{h_y}| \geq 5$, and $|\frac{k\delta_z}{h_z}| \geq 5$
(\citeannex{wand1994fast} suggest $4$ as a ``safe choice''). This means
that many of the terms in $\bm{K}$ can be set to zero.

Next, we define
\begin{IEEEeqnarray}{rCl}
  \lambda_x & = & \min(n_x - 1, \left \lfloor \frac{5h_x}{\delta_x} \right \rfloor)\\
  \lambda_y & = & \min(n_y - 1, \left \lfloor \frac{5h_y}{\delta_y} \right \rfloor)\\
  \lambda_z & = & \min(n_z - 1, \left \lfloor \frac{5h_z}{\delta_z} \right \rfloor)
\end{IEEEeqnarray}
$\lambda_x$, $\lambda_y$ and $\lambda_z$ are the indices beyond which the kernel becomes
practically null (or the dimensions of the image if these indices are
superior to the image size).
We also define
\begin{IEEEeqnarray}{rCl}
  \theta_x & = & 2^{ \left\lceil \log_2(n_x + \lambda_x + 1) \right\rceil } \\
  \theta_y & = & 2^{ \left\lceil \log_2(n_y + \lambda_y + 1) \right\rceil } \\
  \theta_z & = & 2^{ \left\lceil \log_2(n_z + \lambda_z + 1) \right\rceil }
\end{IEEEeqnarray}
$\theta_x$ is the smaller power of 2 that is bigger than $n_x + \lambda_x + 1$\\
Then the non-zero terms of $\bm{K}$ are symmetrized, and $\bm{K}$ is padded with
zeros in the middle to size $(\theta_x \times \theta_y \times \theta_z)$. The reason
for this padding is that the \ac{FFT} is faster if the dimensions of
$\bm{K}$ are highly composite numbers, such as powers of 2. The new matrix
$\bm{K}$ is therefore a matrix of size $(\theta_x \times \theta_y \times \theta_z)$ such
that:
\begin{IEEEeqnarray*}{rCll}
  \bm{K}_{i,j,k}& = & \phi_{\bm{H}}(i\delta_x, j\delta_y, k\delta_z) ,& i=0,\dots,\lambda_x, j=0,\dots,\lambda_y, k=0,\dots,\lambda_z\\
  \bm{K}_{i,j,k}& = & \phi_{\bm{H}}((\theta_x - i)\delta_x, j\delta_y, k\delta_z) ,& i=\theta_x - \lambda_x,\dots,\theta_x - 1, j=0,\dots,\lambda_y, k=0,\dots,\lambda_z\\
  \bm{K}_{i,j,k}& = & \phi_{\bm{H}}(i\delta_x, (\theta_y - j)\delta_y, k\delta_z) ,& i=0,\dots,\lambda_x,j=\theta_y - \lambda_y,\dots,\theta_y-1, k=0,\dots,\lambda_z\\
  \bm{K}_{i,j,k}& = & \phi_{\bm{H}}(i\delta_x,  j\delta_y, (\theta_z - k)\delta_z) ,& i=0,\dots,\lambda_x, , j=0,\dots,\lambda_y, k=\theta_z - \lambda_z,\dots,\theta_z-1\\
  \bm{K}_{i,j,k}& = & \phi_{\bm{H}}(i\delta_x, (\theta_y - j)\delta_y, (\theta_z - k)\delta_z) ,& i=0,\dots,\lambda_x,j=\theta_y - \lambda_y,\dots,\theta_y-1, \\
  &&&k=\theta_z - \lambda_z,\dots,\theta_z-1\\
  \bm{K}_{i,j,k}& = & \phi_{\bm{H}}((\theta_x - i)\delta_x, j\delta_y, (\theta_z - k)\delta_z) ,& i=\theta_x - \lambda_x,\dots,\theta_x - 1, j=0,\dots,\lambda_y,\\ &&&k=\theta_z - \lambda_z,\dots,\theta_z-1\\
  \bm{K}_{i,j,k}& = & \phi_{\bm{H}}((\theta_x - i)\delta_x, (\theta_y - j)\delta_y, k\delta_z) ,& i=\theta_x - \lambda_x,\dots,\theta_x - 1,j=\theta_y - \lambda_y,\dots,\theta_y-1,\\
  &&&k=0,\dots,\lambda_z\\
  \bm{K}_{i,j,k}& = & \phi_{\bm{H}}((\theta_x - i)\delta_x, (\theta_y - j)\delta_y, (\theta_z - k)\delta_z) ,& i=\theta_x - \lambda_x,\dots,\theta_x - 1, \\
  & & &j=\theta_y - \lambda_y,\dots,\theta_y-1, \\
  & & & k=\theta_z - \lambda_z,\dots,\theta_z-1 \\
\end{IEEEeqnarray*}
For example, the first slice (:,:,0) of $\bm{K}$ looks like this:
\begin{equation}
  \bm{K}_{:,:,0} = \begin{bmatrix}
    \phi_{0,0,0}&\phi_{0,1,0}&\dots&\phi_{0,\lambda_y,0}&0&\dots&0&\phi_{0,\lambda_y,0}&\dots&\phi_{0,1,0}\\
    \phi_{1,0,0}&\phi_{1,1,0}&\dots&\phi_{1,\lambda_y,0}&0&\dots&0&\phi_{1,\lambda_y,0}&\dots&\phi_{1,1,0}\\
    \vdots&\vdots&&\vdots&\vdots&&\vdots&\vdots&&\vdots\\
    \phi_{\lambda_x,0,0}&\phi_{\lambda_x,1,0}&\dots&\phi_{\lambda_x,\lambda_y,0}&0&\dots&0&\phi_{\lambda_x,\lambda_y,0}&\dots&\phi_{\lambda_x,1,0}\\
    0&0&\dots&0&0&\dots&0&0&\dots&0\\
    \vdots&\vdots&&\vdots&\vdots&&\vdots&\vdots&&\vdots\\
    0&0&\dots&0&0&\dots&0&0&\dots&0\\
    \phi_{\lambda_x,0,0}&\phi_{\lambda_x,1,0}&\dots&\phi_{\lambda_x,\lambda_y,0}&0&\dots&0&\phi_{\lambda_x,\lambda_y,0}&\dots&\phi_{\lambda_x,1,0}\\
    \vdots&\vdots&&\vdots&\vdots&&\vdots&\vdots&&\vdots\\
    \phi_{1,0,0}&\phi_{1,1,0}&\dots&\phi_{1,\lambda_y,0}&0&\dots&0&\phi_{1,\lambda_y,0}&\dots&\phi_{1,1,0}
  \end{bmatrix}
\end{equation}
Where we have written
$\phi_{i,j,k} = \phi_{\bm{H}}(i\delta_x, j\delta_y, k\delta_z)$ for brevity.
The weights' matrix $\bm{W}$ is then padded with zeros to be the same size
as $\bm{K}$: $\bm{W}$ is of size $(\theta_x, \theta_y, \theta_z)$ and
\begin{IEEEeqnarray}{rCl}
  \bm{W}_{i,j,k} & = \begin{cases} w_{i,j,k} &\text{ if } i < n_x, j < n_y, k < n_z\\
   0 & \text{otherwise}\end{cases}
\end{IEEEeqnarray}

Then the upper corner of the convolution of $\bm{K}$ and $\bm{W}$ gives us the
estimates of our \ac{pdf} at every voxel:

\begin{equation}
  \hat{p}(I) = (\bm{K} * \bm{W})[:n_x, :n_y, :n_z]
\end{equation}
where $*$ is the convolution operator and $I$ is the image:

\begin{equation}
  I = \{(i\delta_x, j\delta_y, k\delta_z), 0 \leq i < n_x, 0  \leq j < n_y, 0 \leq k < n_z\}
\end{equation}

Since the dimensions of $\bm{K}$ and $\bm{W}$ are highly composite
numbers, computing their Fourier Transforms is very fast, and the
convolution is obtained as the inverse Fourier Transforms of the
element-by-element product of the transforms of $\bm{K}$ and $\bm{W}$, according
to the discrete convolution theorem:

\begin{equation}
  \hat{f}(\text{image}) = \mathcal{F}^{-1}(\mathcal{F}(\bm{K}) \odot \mathcal{F}(\bm{W}))
\end{equation}
Where $\mathcal{F}$ is the Fourier transform and $\odot$ is the
Hadamard (element-wise) product.

When using \ac{KDE}, two important decisions are the choice of the
kernel and the choice of the smoothing factor $h$, also called
bandwidth. Theoretical results as well as cross-validation can be used
to make these choices \citeannex{silverman1986density}. We chose a Gaussian
kernel for simplicity and because it is popular for smoothing
MRI images. If we choose
a bandwidth matrix with diagonal $(h^2, h^2, h^2)$, the \ac{FWHM} of
the kernel is equal to $2 \delta h \sqrt{2 \ln(2)}$, where $\delta$ is
the voxel size. We compared $h=0.5$, $h=1$ and $h=2$, and the bandwidth we chose
is $h=1$, which yields a \ac{FWHM} of around \SI{9}{mm}.

\bibliographystyleannex{splncs}
\bibliographyannex{miccai_biblio.bib}

\end{document}